\documentclass[useAMS,usenatbib,usegraphicx]{mn2e}

\usepackage{amssymb}
\usepackage{amsmath}
\usepackage{graphicx} 
\usepackage{epsfig}
\usepackage{color}
\usepackage{bm}
\usepackage{ulem}

\author[Q. Xia et al.]{Qiran Xia$^{1,2}$\thanks{qiranxia@gmail.com}, Chao Liu$^{1,3}$, Shude Mao$^{4,1,5}$, Yingyi Song$^{6}$, Lan Zhang$^{3,7}$, R. J. Long$^{1,5}$,\newauthor Yong Zhang$^{8}$, Yonghui Hou$^{8}$, Yuefei Wang$^{8}$, Yue Wu$^{3}$ \\
$^1$ National Astronomical Observatories, CAS, 20A Datun Road, Chaoyang District, 100012, Beijing, China\\
$^2$ University of Chinese Academy of Sciences, Beijing 100049, China\\
$^3$ Key Laboratory of Optical Astronomy, National Astronomical Observatories, CAS, 20A Datun Road, Chaoyang District, 100012, Beijing, China\\
$^4$ Center for Astrophysics , Department of Physics, Tsinghua University, 10086, Beijing, China\\
$^5$ Jodrell Bank Centre for Astrophysics, School of Physics and Astronomy, The University of Manchester, Oxford Road, Manchester M13 9PL, UK\\
$^6$ Department of Astronomy, University of Michigan, 311 West Hall, 1085 South University Avenue, Ann Arbor, MI 48109, USA\\
$^7$ Departamento de Astronomoa\'{i}de Chile, Camino EI observatorio\#1515, Las Condes, Santiago, Chile\\
$^8$ Nanjing Institute of Astronomical Optics \& Technology, National Astronomical Observatories, Chinese Academy of Sciences, Nanjing 210042, China\\}

\title{Determining the local dark matter density with LAMOST data}

\begin{document}

\date{Accepted ...... Received ...... ; in original form......   }

\pagerange{\pageref{firstpage}--\pageref{lastpage}} \pubyear{2015}

\maketitle
\label{firstpage}

\begin{abstract}
Measurement of the local dark matter density plays an important role in both Galactic dynamics and dark matter direct detection experiments. However, the estimated values from previous works are far from agreeing with each other. In this work, we provide a well-defined observed sample with 1427 G \& K type main-sequence stars from the LAMOST spectroscopic survey, taking into account selection effects, volume completeness, and the stellar populations. We apply a vertical Jeans equation method containing a single exponential stellar disk, a razor thin gas disk, and a constant dark matter density distribution to the sample, and obtain a total surface mass density of $78.7 ^{+3.9}_{-4.7}\,\rm M_{\odot}\,\rm pc^{-2}$ up to 1\,kpc and a local dark matter density of $0.0159^{+0.0047}_{-0.0057}\,\rm M_{\odot}\,\rm pc^{-3}$. We find that the sampling density (i.e. number of stars per unit volume) of the spectroscopic data contributes to about two-thirds of the uncertainty in the estimated values. We discuss the effect of the tilt term in the Jeans equation and find it has little impact on our measurement. Other issues, such as a non-equilibrium component due to perturbations and contamination by the thick disk population, are also discussed.

\end{abstract}


\section{Introduction}\label{sect:intro}

\begin{figure}
	\centering
 	\includegraphics[width=3in]{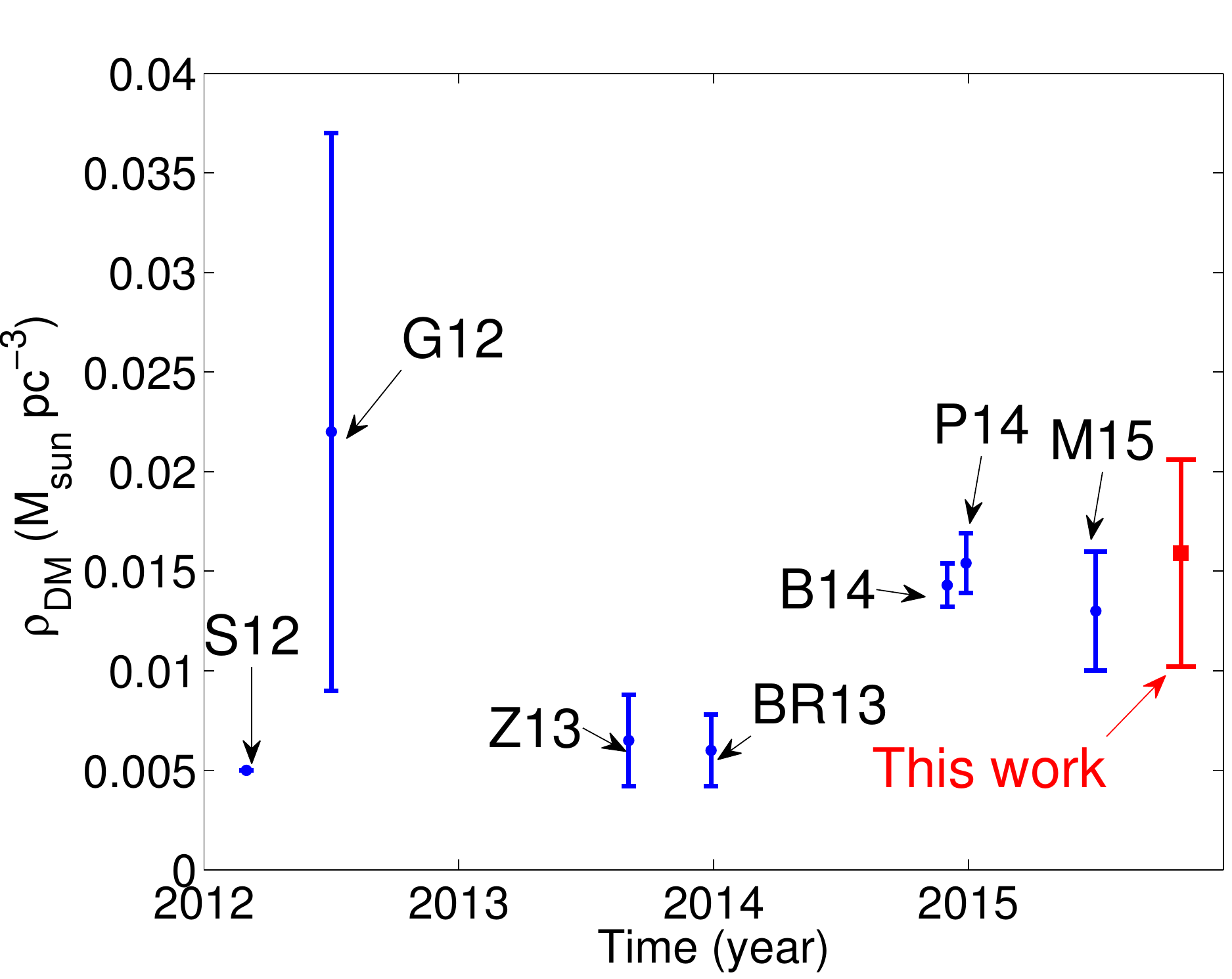} 
 	\caption{Comparison of the measurements of the local dark matter density in the past three years. The points are from Smith et al. (2012; S12), Garbari et al. (2012; G12), Zhang et al. (2013; Z13), Bovy \& Rix (2013; BR13), Piffl et al. (2014; PB14), Bienaym{\'e} et al. (2014; BF14), and McKee et al. (2015; M15). Our result is shown as the red square with error bar.}
	\label{fig:rr}
\end{figure}

The local dark matter density is an important quantity not only because it can anchor the total dark matter mass of the Galactic halo, but also because it gives strong constraints in the search for dark matter particles in ground-based laboratories~\citep{Read14}. Since~\citet{Oort32,Oort60}, this value has been measured in many works with different models and observational samples (e.g. \citealt{KG89a,KG89b,KG89c,KG91,HF00,HF04,WB10}). However, these results are far from agreeing with each other. Fig.~\ref{fig:rr} demonstrates the discrepant results, most of which are from recent survey data in the past three years (\citealt{Smith12,Garbari12,Zhang13,BR13,Bienayme14,Piffl14,McKee15}). Further comparisons with historic works are displayed in~\citet{Read14}.
The discrepancies may be due to two factors: 1) different assumptions, simplifications, and dynamical modelling methods have been used in the different works and 2) selection effects, systematics in the parameterizations, and the analysis methods for observed data differ greatly across the various works. In this sense, results cannot be compared naively unless differences in the modeling and observations are well understood.

In general, modelling techniques can be divided into two types: one is to constrain the local dark matter density with the vertical kinematics of stars in the solar neighborhood (e.g.~\citealt{KG89a,KG89b,KG89c,HF00, HF04,Garbari12,Zhang13}), while the other utilises measurements of the rotation curve (e.g. \citealt{DB98,Fich89,Merrifield92,Sofue09,WB10,CU10,McMillan11,Piffl14}). 
Even only in the first category, different model assumptions may lead to quite different results. For instance, \citet{Garbari11} carefully compared the minimum assumption method developed by themselves with the Holmberg \& Flynn method~\citep{HF00} using data from an N-body simulation. They found that the assumed form of the distribution function in the Holmberg \& Flynn method may be violated when the Galactic disk is not isothermal and axisymmetric, or when the observed data is beyond about 1 scale height. 
Therefore, systematic comparisons between various models are necessary so that the assumptions can be assessed.
However, model comparison can not be properly made unless the observed data are carefully chosen with a good understanding of the uncertainties in the observables (stellar parameters, distance, and kinematics) and the selection effects. 

The observed data used in previous works are from quite different sources with different systematics in parameter determination and selection effects. For example, the distances from the Sloan Digital Sky Survey (hereafter SDSS) estimated by \citet{An09} are ~9\% smaller than those of \citet{Ivezic08}, according to \citet{Zhang13}. The sky coverage of SDSS, a northern sky survey, is quite different from the Radial Velocity Experiment (hereafter RAVE), a southern sky survey. Consequently, proper motions used in RAVE-based measurements (\citealt{Bienayme14, Piffl14}) may have unknown systematics compared with those provided by SDSS, which are used by~\citet{Zhang13}, due to the lack of uniform calibrations. The stellar tracers used are also different, e.g. \citet{Zhang13} used K dwarfs while \citet{Bienayme14} used red clump stars. All these differences can significantly affect the results. 

The aim of this work is to provide an observed dataset with well understood selection effects within a proper spatial volume which from to estimate the dark matter density in the solar neighborhood. Such a dataset allows us to validate the assumptions adopted by different models based on vertical stellar kinematics. In the first of a series of works, we first estimate the local dark matter density using the vertical Jeans equation method on a well defined dataset from the Large Sky Area Multi-Object Fibre Spectroscopic Telescope (hereafter LAMOST) survey. 

The on-going LAMOST survey \citep{Zhao12} is observing spectra for several millions of stars in low resolution and will provide the largest spectroscopic sample within a few kpc around the Sun in the next few years \citep{Deng12}. We construct the observed sample from the LAMOST data carefully taking account of selection effects, volume completeness, reliable distance estimates, and sufficiently accurate vertical velocities. 

The vertical Jeans equation method is one of the classical models and was first used by \citet{KG89a,KG89b,KG89c}. They estimated the surface density of the baryonic matter and that of the total mass to be $48 \pm 8\,\rm M_{\odot}\,{\rm pc}^{-2}$ and $71 \pm 6\,\rm M_{\odot}\,{\rm pc}^{-2}$, respectively, within a vertical distance of $1.1\ \rm{kpc}$ from the Galactic mid-plane near the Sun. Recently, \citet{Garbari12} constrained the parameters of their improved vertical Jeans equation method using a Markov Chain Monte Carlo (hereafter MCMC) technique and obtained a local dark matter density of $\rho_{\rm DM} = 0.025^{+0.014}_{-0.013}\,\rm M_{\odot}\,{\rm pc}^{-3}$ using the same data as Kuijken \& Gilmore (1989b, hereafter KG). \citet{Zhang13} applied a similar method to K dwarf stars from SDSS DR8 and derived a local dark matter density of $\rho_{\rm DM} = 0.0065\pm0.0023\,\rm M_{\odot}\,{\rm pc}^{-3}$.

The paper is organized as follows. In section 2, we first choose the data sample and then correct for selection effects. 
 In section 3, we detail the vertical Jeans equation method suitable for the dataset.
The results and discussion are shown in section 4 . Finally, we conclude in section 5.

\section{Data}\label{sect:data}

LAMOST (also called the Guoshoujing telescope) is a 4 metre quasi-meridian reflective Schmidt telescope being used for a spectroscopic survey \citep{Cui12,Zhao12}. The survey with millions of stellar spectra enables a broad range of studies to be undertaken on the Milky Way \citep{Deng12}. The LAMOST DR2 catalogue contains 4,136,482 spectra,  of which 1,085,404 spectra have atmospheric parameter estimates \citep{Wu14}. Distances are estimated from a Bayesian approach~\citep{Carlin15} with uncertainties of about 20\%.

This large dataset covers the solar neighborhood well and thus provides an ideal database for the selection of the data sample for the measurement of the local dark matter density. First, we select about 4000 stars located with $b>85^\circ$, i.e. close to the north Galactic pole, so that the radial velocities estimated from the spectra are approximately equal to the vertical velocities (a more detailed discussion is in section~\ref{veldisp}). Hence, the vertical kinematics of these stars avoid proper motions, which are far less reliable than the radial velocities. Second, we select stars of a specific spectral type as the tracer population. It is noted that main-sequence stars with $\rm T_{eff}>6000\ K$ are dominated by young stars with age $<4$\,Gyr and thus may not be as relaxed as old stars~\citep{tian15}. Therefore we only select stars with $4000<\rm T_{eff}<6000$\,K to ensure that most of the tracers are already in equilibrium. Stars with $\rm T_{eff} < 4000\ K$ are not selected because atmospheric parameters cannot be accurately estimated due to strong molecular bands in the spectra. 

Dynamical modelling requires stars from both low and high $z$ (here $z$ represent the vertical height above the equatorial plane) regimes in order to detect both the baryonic component and the dark matter. Consequently, we select stars located between 200 and 1500\,pc in $z$. However, not all types of spectroscopically observed stars are volume complete in this range of $z$ because of the limiting magnitude, $r=17.8$\,mag, for the survey. Stars with J band absolute magnitudes between 3 and 4.5\,mag are selected as they are volume complete in the vertical range from 200\,pc to 1500\,pc.

\begin{figure}
	\centering
 	\includegraphics[width=3in]{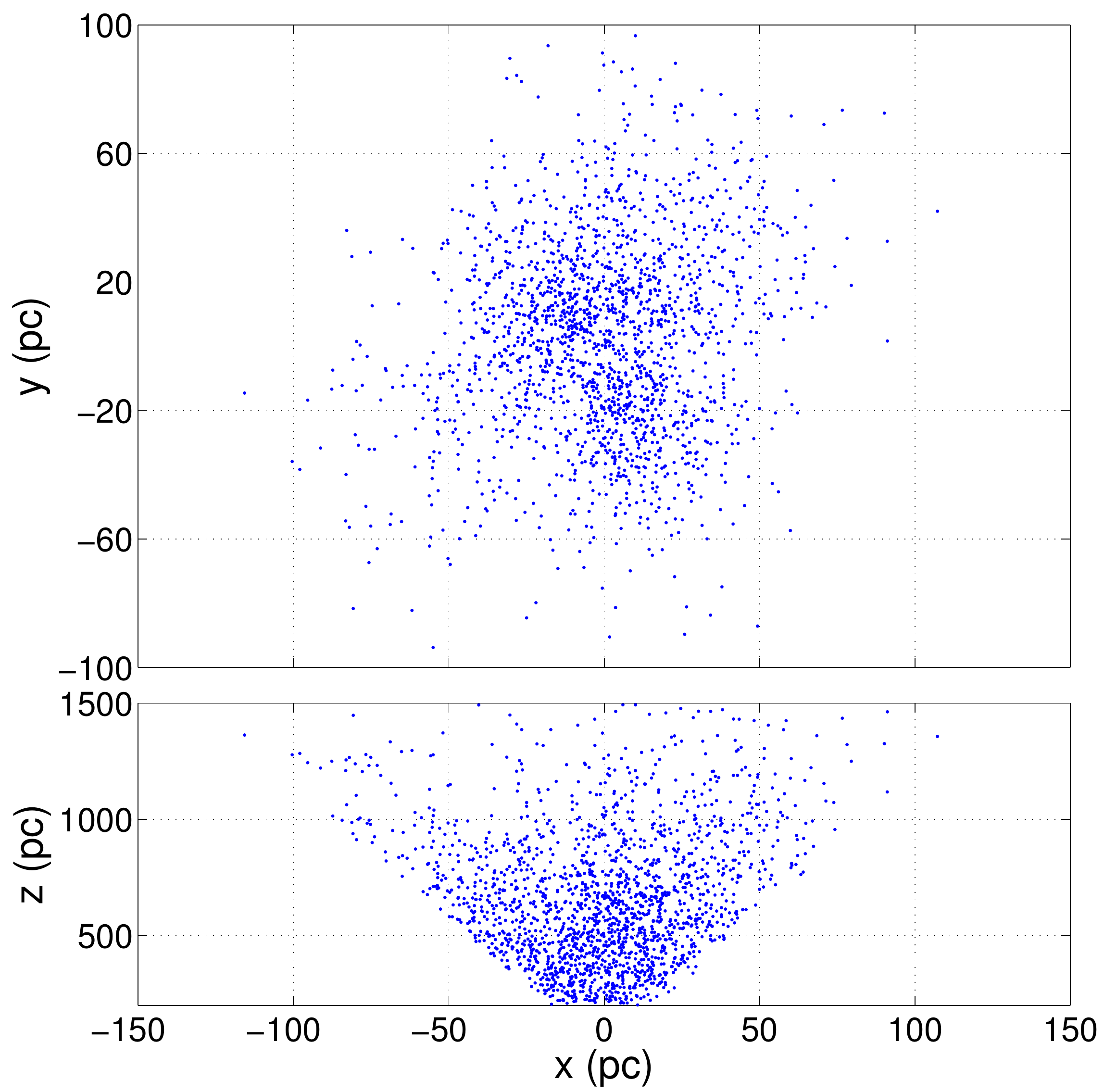} 
 	\caption{The positions of the observed sample used in this work in heliocentric Cartesian coordinates. The positive $x$ is toward the Galactic centre, and the positive $z$ is toward the north Galactic pole.}
	\label{fig:location}
\end{figure}

It is noted that multiple populations with different scale heights in the dataset may increase the uncertainty and induce systematics in the measurement of dark matter density \citep{Hessman15}. We therefore only select stars in the thin disk  by requiring [Fe/H]$>-0.5$\,dex to keep the density profile to be a single exponential. This selection effectively removes the thick disk population (see the detailed discussion in section~\ref{sect:thickdisc}).
Finally, only stars with S/N $>10$ are selected to ensure the accuracy of their radial velocities. The final sample contains 1427 stars and Fig.~\ref{fig:location} shows their 3D positions in heliocentric Cartesian coordinates. 

\subsection{Selection effects and stellar density}

\begin{figure}
	\centering
 	\includegraphics[width=3in]{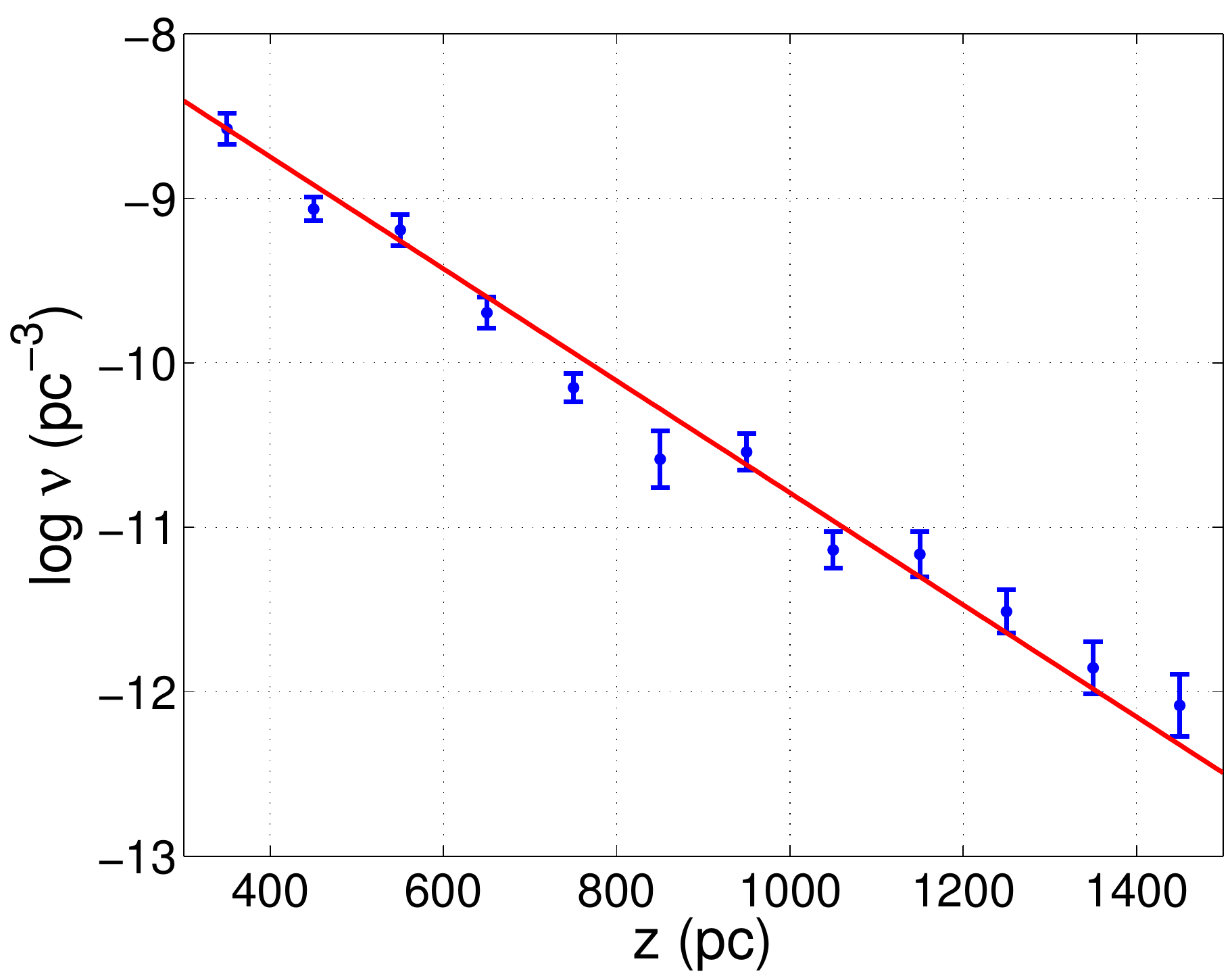} 
 	\caption{The dots with error bars are the vertical stellar density for our observational sample. The red solid line is the best fit exponential density profile model.}
	\label{fig:ndf}
\end{figure}

In general, sampling from a spectroscopic survey is significantly affected by the survey's selection strategy, observational conditions, data reduction and so on. These induce substantial sample selection effects which must be taken into account in the determination of the stellar density. 

For the LAMOST survey, although the field of view covers 20 square degrees, only fewer than 3000 stars within the field can be observed (the remaining $\sim1000$ fibers are used for galaxies, special objects, flux standards, sky background and so on.). According to~\citet{Carlin12}, the selection function in the LAMOST survey is simple. Indeed, no selection bias is found in color indices in the survey by comparing the DR2 catalogue with the corresponding input catalogue. The only selection effect is in magnitudes because 1) bright stars are easier to observe than fainter ones; and 2) the stellar density of brighter stars is much lower than that of the fainter ones, thus the sampling rate for bright stars is higher than that for the fainter ones.

Therefore we assume that for a given line of sight and a sufficiently small region in the color--magnitude (CM) plane, LAMOST observed stars are a subsample selected from a photometric catalogue, which is taken to be complete. In this work, we use the 2MASS catalogue as the photometric dataset~\citep{2mass}, since it covers more than 90\% of the LAMOST data. The selection function for a given line of sight, $\theta$, and a small CM bin can be written as 
\begin{equation}\label{eq16}
{f_{{\rm 2MASS},\,\theta,\,{\rm CM}}(z)} = \varpi _{\theta,\,{\rm CM}}\ {f_{{\rm LAMOST},\,\theta,\,{\rm CM}}(z)}, 
\end{equation}
where $z$ is the vertical distance, $f_{{\rm 2MASS},\,\theta,\,{\rm CM}}(z)$ the stellar density of 2MASS at $z$ and is unknown,  $f_{{\rm LAMOST},\,\theta,\,{\rm CM}}(z)$ is the stellar density of the LAMOST data at $z$ and can be estimated from the spectroscopic sample. $\varpi_{\theta,\,{\rm CM}}$ in eq.~(\ref{eq16}) is the ratio of the number of stars from the photometric catalogue to that from the spectroscopic data within $\theta$ and CM. The total stellar density at $z$ should be
\begin{equation}
\begin{aligned}
\nu_{obs}(z) &= \sum_{{\rm CM},\,\theta}f_{{\rm 2MASS},\,\theta,\,{\rm CM}}(z)\\
 &= \sum_{{\rm CM},\,\theta}{\varpi _{\theta,\,{\rm CM}}\ f_{{\rm LAMOST},\,\theta,\,{\rm CM}}(z)}.
 \end{aligned}
\end{equation}

The dots with error bars in Fig.~\ref{fig:ndf} show the stellar density profile with bin size of 100 pc. Errors are estimated from bootstrap resampling. It is clear that the stellar density profile is approximately a straight line in log\,$\rm \nu$ vs. $z$ which implies that the majority of stars in our sample belong to the thin disk with a single scale height.

\subsection{Velocity dispersion}\label{veldisp}
The uncertainty in the line-of-sight velocities from the LAMOST spectra is around $5\,\rm km\,s^{-1}$~\citep{Gao14}. Since the Galactic latitude of the sample stars is larger than 85$^\circ$, the line-of-sight velocity can be approximately used as the vertical velocity. Thus, we avoid the large uncertainty from proper motions, which suffer from significant systematic biases and have uncertainties that increase with the distance. 

\begin{figure}
	\centering
 	\includegraphics[width=3in]{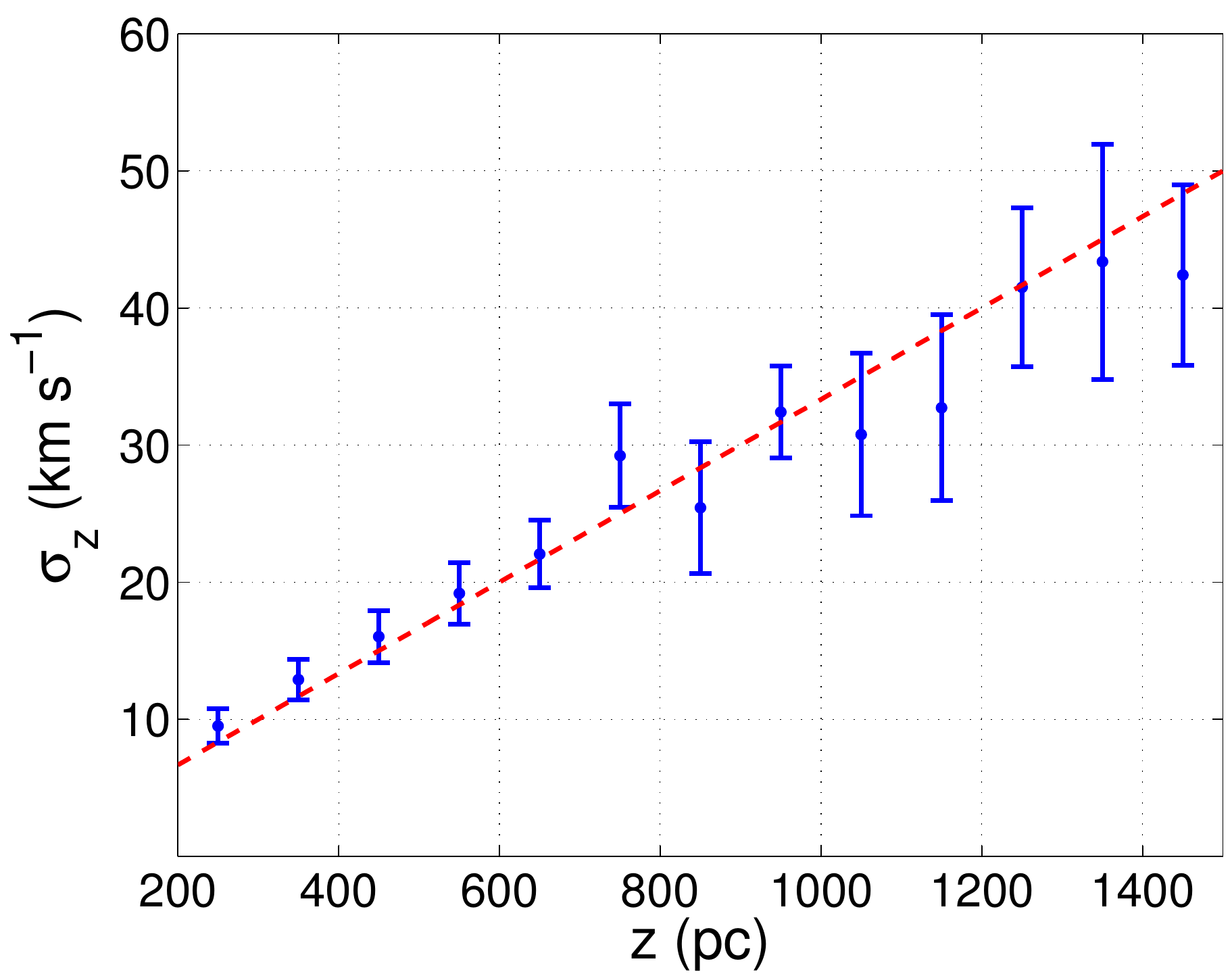} 
 	\caption{The result of the Monte Carlo simulation for the verification of the replacement of the vertical velocities with the line-of-sight velocities. The red dashed line stands for the input vertical velocity dispersions. The blue dots with error bars show the vertical velocity dispersions recovered from the line-of-sight velocities.}
	\label{fig:tv}
\end{figure}

In order to verify that the difference between the line-of-sight velocity and the vertical velocity is negligible, we run a Monte Carlo simulation using a simple but well-motivated model for the Milky Way disk in the same range of the Galactic latitude as the observed sample:
\begin{equation}
\begin{aligned}
&\nu(z) \propto {\rm exp}(-\frac{z}{h}),\ h=300\,{\rm pc},\\
&\sigma_{z}(z) = v_{0}\times\frac{z}{z_{0}},\ v_{0}=\frac{10}{3}\,{\rm km\,s^{-1}},\ z_{0}=100\,{\rm pc},\\
&\sigma_{r} = \sigma_{\theta} = 2\sigma_{z},\\
&\langle v_{r} \rangle =\langle v_{z} \rangle = 0\,{\rm km\,s^{-1}},\\
&\langle v_{\theta} \rangle= 220\,{\rm km\,s^{-1}},
\end{aligned}
\end{equation}
where $\nu(z)$ is the vertical stellar density profile, $\langle v_{i} \rangle$ and $\sigma_{i}$ are the mean velocity and velocity dispersion respectively in three dimensions in the cylindrical coordinates. The simulation result is shown in Fig.~\ref{fig:tv}. The estimated line-of-sight velocity dispersions at are consistent with the red dashed line, which gives the input model vertical velocity dispersions. 

\begin{figure}
	\centering
 	\includegraphics[width=3in]{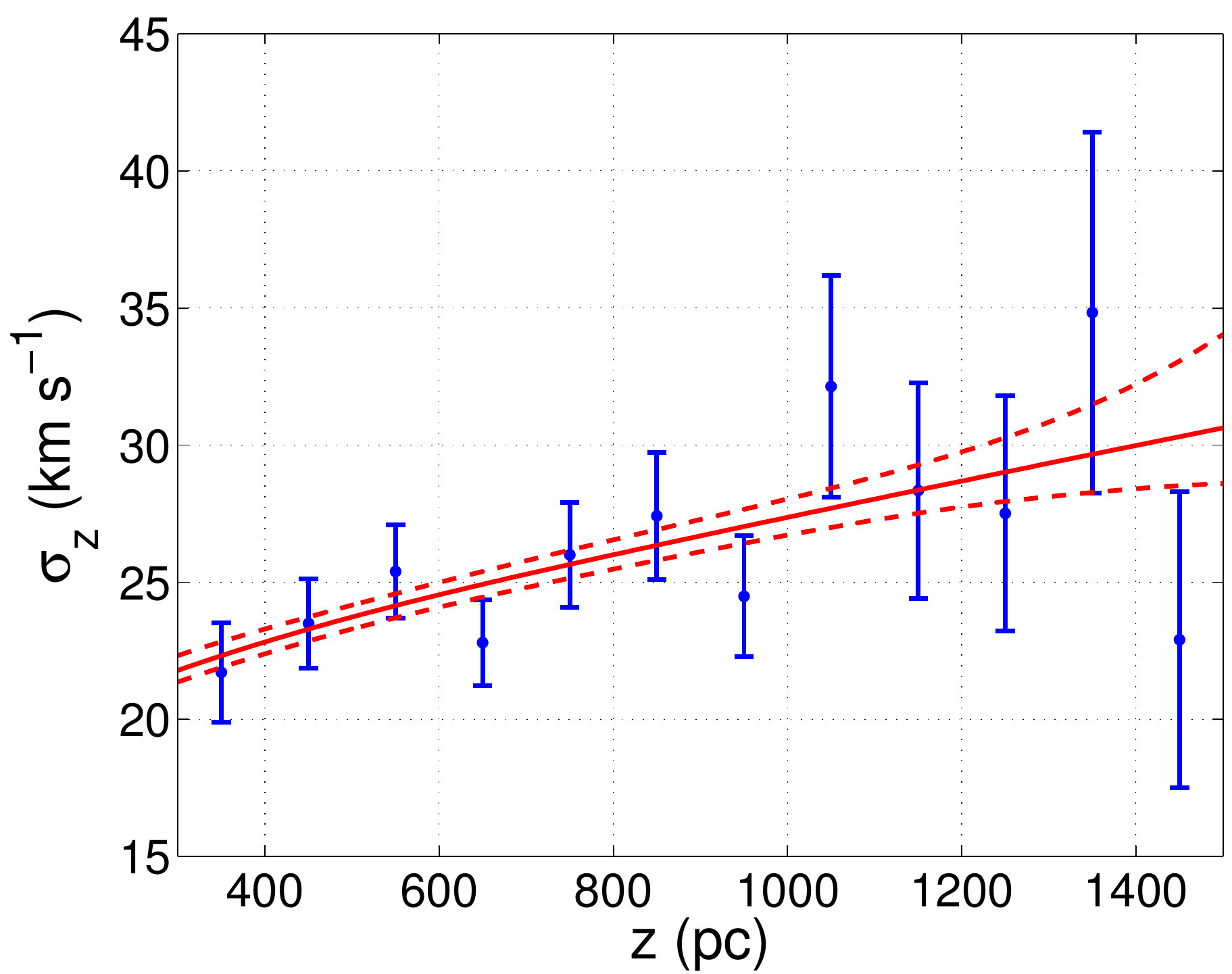} 
 	\caption{The dots with error bars represent the vertical velocity dispersion for our observational sample. The red solid line is the best fit of the velocity dispersion and the two red dashed lines show the 1-$\sigma$ region of the PDF for velocity dispersion.}
	\label{fig:vdf}
\end{figure}

The dots with error bars in Fig.~\ref{fig:vdf} show the vertical velocity dispersions derived from the line-of-sight velocities of our observed sample. The velocity dispersions are derived from the standard deviations of the velocities and the measurement errors of the velocities have been removed by deconvolution assuming Gaussian errors.

It is worth pointing out that the binning in $z$ used in Figs.~\ref{fig:ndf} and~\ref{fig:vdf} is just for plotting. The fitting procedure we employ for our model parameters uses the spatial and kinematic information for individual stars without binning (see the logarithmic likelihood function eq.~\ref{eq14} in section~\ref{sect:Kz}).

\section{Method}\label{sect:method}


In this work, we focus on the vertical Jeans equation method following \citet{Garbari12} and \citet{Zhang13}. The first three assumptions of the method are as follows 
\begin{enumerate}
\item[1)] the Milky Way is axisymmetric;
\item[2)] the stars used for tracing the gravitational potential are individually in dynamical equilibrium;
\item[3)] the tilt term in the Jeans equation below (eq. \ref{eq1}) is negligible.
\end{enumerate}

\subsection{Vertical Jeans equation}
For an axisymmetric system, the vertical part of the Jeans equation in cylindrical coordinates is
\begin{equation} \label{eq1}
\frac{\partial}{\partial{z}} (\nu \sigma^{2}_{zz}) + \frac{1}{R\ \nu} \frac{\partial}{\partial{R}} (R \nu \sigma^{2}_{Rz})= - \nu \frac{\partial{\Phi}}{\partial{z}},
\end{equation}
where $\sigma^{2}_{ij} = \overline{v_{i}v_{j}} - \bar{v}_{i}\bar{v}_{j}$ is the velocity dispersion tensor \citep{BT08}, $\Phi$ the Galactic potential, $\nu$ the vertical stellar density profile of the tracers. The density profile, $\nu$, the velocity dispersion, $\sigma^{2}_{z}$, and the Galactic potential, $\Phi$, in eq. (\ref{eq1}) only depend on the vertical distance $z$. Here we assume that the tilt term, $\frac{1}{R} \frac{\partial}{\partial{R}} (R \nu \sigma^{2}_{Rz})$, is negligible since we only consider a small volume, e.g. an cone with width of 0.1\,kpc and thickness of 1.5\,kpc around the Sun (\citealt{Garbari12,Zhang13,Read14}). Then eq. (\ref{eq1}) can be simplified as
\begin{equation}\label{eq2}
\frac{d}{d z} [\nu(z) \sigma^{2}_{z}(z)] = - \nu {\frac{d \Phi (z)}{d z}}\Big |_{R_{0}},
\end{equation}
where $\Phi(z)|_{R_{0}}$ is the vertical gravitational potential in the solar vicinity, $R_{0}$ is the Sun's distance to the Galactic centre. Further discussion about whether the tilt term can be neglected is in section~\ref{sect:tiltterm}. 

\subsection{One-dimensional Poisson equation}
In one dimension, the vertical mass distribution is connected with the Galactic potential via the following one-dimensional Poisson equation:
\begin{equation}\label{eq3}
4\pi G \rho_{\rm tot}(z)|_{R_{0}} = \frac{d^{2}\Phi (z)}{d^{2}z}\Big |_{R_{0}}.
\end{equation}
We assume that
\begin{enumerate}
\item[4)] the Galactic potential comprises baryonic matter (stars and gas) and dark matter. 
\end{enumerate}
Thus,
\begin{equation}\label{eq4}
\Phi (z) = \Phi_{\rm disk} (z) + \Phi_{\rm gas} (z) + \Phi_{\rm DM} (z),
\end{equation}
where $\Phi_{\rm disk} (z)$, $\Phi_{\rm gas} (z)$, and $\Phi_{\rm DM} (z)$ are the potentials of the stellar disk, the gas disk, and the dark matter, respectively.
Similarly, 
\begin{equation}\label{eq5}
\rho_{\rm tot}(z) = \rho_{\rm disk}(z) + \rho_{\rm gas} (z) + \rho^{eff}_{\rm DM}(z),
\end{equation}
where $\rho_{\rm tot}(z)$, $\rho_{\rm disk}(z)$, $\rho_{\rm gas}(z)$, and $\rho^{eff}_{\rm DM}(z)$ are the densities for the total mass, the stellar disk, the gas disk, and the effective dark matter including the circular velocity term, respectively. $\rho^{eff}_{\rm DM}(z)$ can be written as \citep{Garbari11}
\begin{equation}\label{eq6}
\rho^{eff}_{\rm DM}(z)= \rho_{\rm DM}(z) - {\frac{1}{4\pi G R} \frac{\partial}{\partial{R}} V^{2}_{c}(R)}\Big|_{R_{0}},
\end{equation}
where, $V^{2}_{c}(R)|_{R_{0}}$ is the circular velocity at the solar circle. Here, we assume that
\begin{enumerate}
\item[5)] in the solar neighborhood, the rotation curve is flat \citep{B12a}.
\end{enumerate}
Under this assumption, the last term on the right hand side of eq. (\ref{eq6}) can be neglected, i.e. $\rho^{eff}_{\rm DM}(z)= \rho_{\rm DM}(z)$. Hereafter, we use $\rho_{\rm DM}(z)$ instead of $\rho^{eff}_{\rm DM}(z)$.


\subsection{Modelling $K_z$}\label{sect:Kz}
The first derivative of the Galactic potential $\Phi(z)$ represents the gravitational force perpendicular to the Galactic plane,
\begin{equation}\label{eq7}
K_{z}(z) \equiv - \frac{d \Phi (z)}{d z}.
\end{equation}
Combining eqs. (\ref{eq2}), (\ref{eq3}), and (\ref{eq7}), the vertical stellar density profile and velocity dispersion of the tracer stars can be related to gravitational force (including the dark matter contribution) by,
\begin{equation}\label{eq8}
\frac{d}{d z} [\nu(z) \sigma^{2}_{z}(z)] = \nu(z) K_{z}(z).
\end{equation} 
Additional assumptions are required to model the mass distribution in the solar neighborhood:
\begin{enumerate}
\item[6)] the vertical mass distribution of stellar disk is exponential with scale hight $\rm z_{h}$;
\item[7)] the gas disk is razor thin, i.e. without thickness;
\item[8)] the dark matter density is a constant in the volume covered by our sample.
\end{enumerate}

With these assumptions, the total vertical mass distribution can be written as
\begin{equation}\label{eq9}
\rho_{\rm tot}(z) = \rho_{\star, 0}\ \rm{exp}\left(-\frac{z}{z_{h}}\right) + \Sigma_{gas,0}\ \delta(z)+ \rho_{\rm DM}.
\end{equation}
Function $K_{z}$ can be obtained by integrating eq.~(\ref{eq7}),
\begin{equation}\label{eq10}
\begin{aligned}
K_{z}(z) &= - \int_{0}^{z} 4\pi G\rho_{\rm tot}(z') dz' \\
&= - 2\pi G \left\{ \Sigma_{\star} \left[ 1- {\rm exp} \left( -\frac{z}{\rm z_{h}} \right) \right] + \Sigma_{\rm gas} + 2\rho_{\rm DM}\,z \right\},
\end{aligned}
\end{equation}
where $\Sigma_{\star}$ is the total surface mass density of the stellar disk, $\Sigma_{\rm gas}$ the surface density of the gas disk, which is taken as $13.2\rm \,M_{\odot}\,pc^{-2}$ \citep{Flynn06}, and $\rho_{\rm DM}$ the local dark matter volume density. At the mid-plane, we have $\frac{d \Phi (z)}{dz}\Big|_{0} = 0$ because of symmetry. According to the observed data, the stellar density of the tracers can be assumed to be a single exponential profile,
\begin{equation}\label{eq11}
\nu(z) = \nu_{0}\ {\rm exp}\left( -\frac{z}{\rm h}\right).
\end{equation}
The four free parameters in the stellar density profile and the $K_{z}$ model is denoted as $\mathbf{p} = (\rm h,\, \Sigma_{\star},\, \rho_{\rm DM},\, z_{h})$.

The velocity dispersion can be solved analytically by integrating eq.~(\ref{eq8}),
\begin{equation}\label{eq12}
\begin{aligned}
&\int_{z_{0}}^{z} d z'\ \frac{d}{d z'} [\nu(z') \sigma^{2}_{z}(z')] = \int_{z_{0}}^{z} d z'\ \nu(z') K_{z}(z'), \\
&\sigma^{2}_{z}(z) = \frac{1}{\nu(z)} \left[\int_{z_{0}}^{z} \nu(z') K_{z}(z')\ dz' - \nu(z_{0})\sigma^{2}_{z}(z_{0})\right], \\
&\sigma^{2}_{z}(z) = f(z) + \frac{\nu(z_{0})\sigma^{2}_{z}(z_{0}) - \nu(z_{0})f(z_{0})}{\nu_{0} \ {\rm exp}\left( -\frac{z}{\rm h}\right)},
\end{aligned}
\end{equation}
where
\begin{equation}\label{eq13}
f(z)= 2\pi G h \left\{ \Sigma_{\star}\left[ 1- {\rm \frac{z_{h}}{h+z_{h}}}{\rm exp}(-\frac{z}{\rm z_{h}}) \right] + \Sigma_{\rm gas} + 2\rho_{\rm DM}(z+\rm h) \right\}
\end{equation}
and $z_{0}$ is the boundary condition for the integration. 
Eqs. (\ref{eq12}) and (\ref{eq13}) imply that the contribution of the baryonic matter increases as a negative exponential with $z$, while the contribution of the dark matter increases linearly. It is known that $\rm h$ is about 300\,pc for the thin disk and since most of the baryonic matter is contributed by the stellar thin disk, $\rm z_{h}$ may be slightly larger because of the thick disk, but cannot be too different from $\rm h$. Thus, the baryonic matter should dominate the low $z$ regime within a few hundred pc above the Sun. The dark matter will take over when $z$ is beyond a few scale heights ($>\sim1$\,kpc). Therefore, in order to accurately measure the contributions from both the baryons and dark matter the observed sample has to cover both the low $z$ ($\sim200$\,pc) and high $z$ ($\sim1500$\,pc) regimes. 

In practice, we find the best-fitting parameters $\mathbf{p}$ using the following steps without binning the data:
\begin{enumerate}
\item[1)] The initial conditions of the model parameters $\mathbf p$ are chosen.
\item[2)] The stellar density profile and the velocity dispersion within the range from 100\,pc to 300\,pc are used as the boundary condition $\nu(z_{0})\sigma^{2}_{z}(z_{0})|_{z_{0} = 200\,\rm pc}$.
\item[3)] The stellar density profile $\nu_{\rm model}(z)$ and the vertical velocity dispersion profile $\sigma_{z,\, {\rm model}}(z)$ are predicted according to eqs. (\ref{eq11}) and (\ref{eq12}) for the given $\mathbf{p}$.
\item[4)] The log-likelihood is given by,
\begin{equation}\label{eq14}
\begin{aligned}
{\rm ln}\ L =& -n\,{\rm ln}(A) - \sum_{i}{\varpi_{i}z_{i}/{\rm h}}\\
&-\sum_{i}{{\rm ln}[\sqrt{2\pi}\sigma_{z,\, {\rm model}}(z_{i})]}\\
&-\frac{1}{2}\sum_{i}{\left[ \frac{v_{i} - m_{v}}{\sigma_{z,\, {\rm model}}(z)} \right]^{2}},
\end{aligned}
\end{equation}
where $i = 1,\cdot\cdot\cdot, n$ is the label of the tracers, $m_{v}$ the mean velocity as a free parameter, and $\varpi_{i}$ the weight of the $i$-th star (see section 3.1).  $A$ is the normalization parameter of the number density, 
\begin{equation}\label{eq15}
A=\frac{1}{\int_{z_{\rm min}}^{z_{\rm max}}{\nu_{0}{\rm exp}(-\frac{z}{\rm h})\,dz}},
\end{equation}
where, $z_{\rm min}$ and $z_{\rm max}$ are the lower and upper limits of the vertical distance of the tracer population, respectively.
\item[5)] Finally, we run MCMC using \textit{emcee}\footnotemark[1] (Foreman-Mackey et al. 2013) to obtain the best fit parameters.
\footnotetext[1]{Emcee is an MIT licensed pure-Python implementation of Goodman \& Weare's Affine Invariant MCMC Ensemble sampler.}\\
\end{enumerate}

\section{Results and DISCUSSIONS}\label{sect:resdis}
\begin{table}
\centering
\caption{The best fit values of the parameters.}\label{tab:results}
\begin{tabular}{clccc}
\hline\hline
\small{Parameter} & \small{Value}\\
\hline
\small{$\Sigma_{\star}\ (\rm M_{\odot}\ \rm pc^{-2})$} & \small{$40.5^{+7.1}_{-6.6}$} \\
\small{$\rm \rho_{\rm DM} \ (M_{\odot}\ \rm pc^{-3})$} & \small{$0.0159^{+0.0047}_{-0.0057}$}\\
\small{$\rm z_{h}\ (pc)$} & \small{$588^{+151}_{-192}$}\\
\small{$\rm h\ (pc)$} & \small{$293.8\pm0.5$}\\
\hline
\end{tabular}
\end{table}

\begin{figure*}
	\centering
 	\includegraphics[width=1\textwidth]{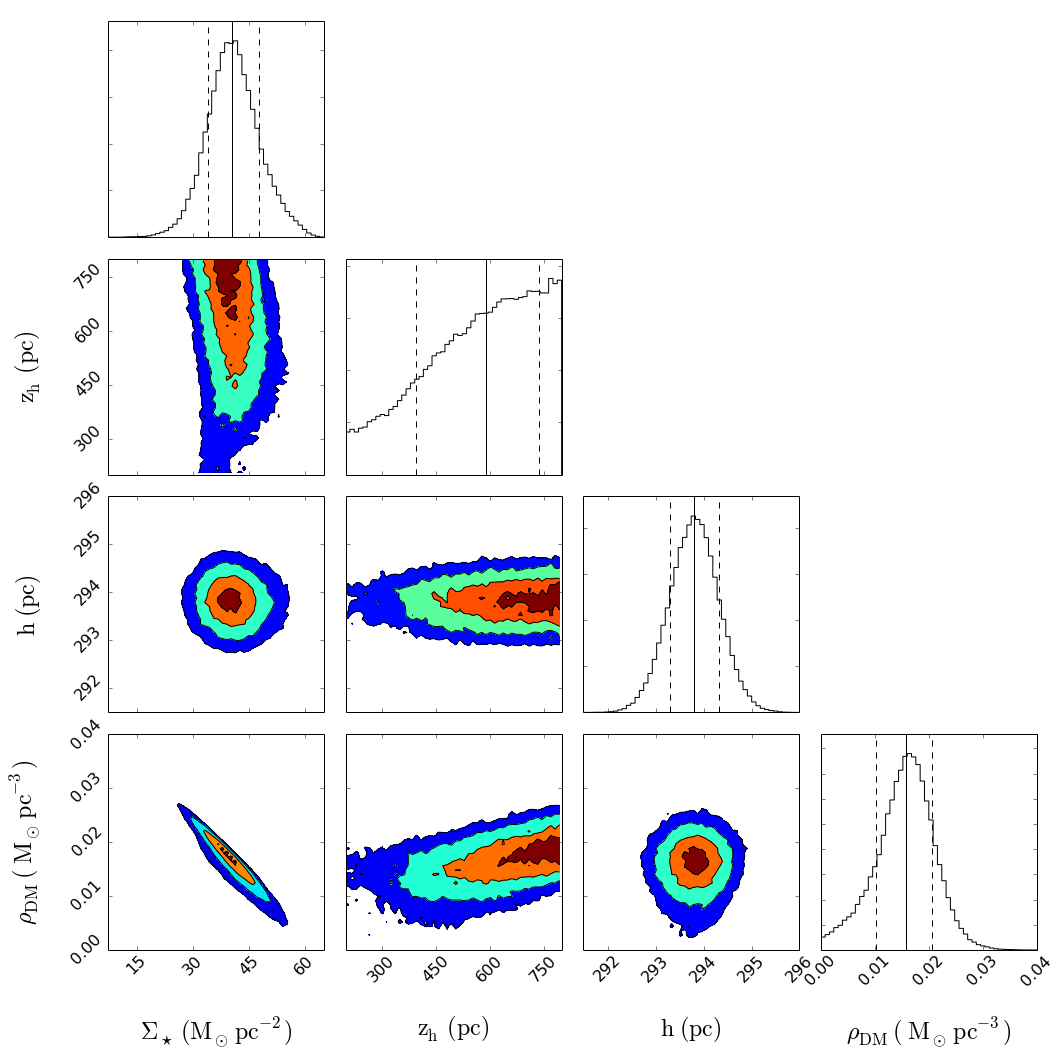} 
 	\caption{The MCMC result of the model parameters. The parameters from the left to the right are $\Sigma_{\star}$, $\rm z_h$, $\rm h$, and $\rho_{\rm DM}$, respectively. The parameters from the top to the bottom are $\rm z_h$, $\rm h$, and $\rho_{\rm DM}$, respectively. The contours display the 0.5, 1, 1.5, and 2 $\sigma$ levels. The solid lines in the histogram panels indicate the median values, and the dashed lines indicate the 1-$\sigma$ regions.}
	\label{fig:pdfs}
\end{figure*}

We apply the model described in section~\ref{sect:method} to the LAMOST sample selected in section~\ref{sect:data}. In the MCMC phase, we use 50 chains each with 50000 steps. The posterior probability density functions (PDFs) of the four parameters are shown in Fig.~\ref{fig:pdfs}. The best-fitting values, i.e. the median values in the MCMC random drawing, are listed in Table 1. The model predicted vertical stellar density and velocity dispersion profiles are over-plotted in Figs.~\ref{fig:ndf}  and~\ref{fig:vdf}, respectively. In Fig.~\ref{fig:vdf}, the  red solid line is the model velocity dispersion and the red dashed lines are the 1-$\sigma$ region. Both vertical profiles are in good agreement with the binned data, although, as previously stated, we use the unbinned data in the model fitting.
The best-fitting value of the scale height of the exponential stellar density profile is $293.8 \pm 0.5$ pc, which is consistent with previous works~\citep{Juric08}. 

\begin{figure}
	\centering
 	\includegraphics[width=3.3in]{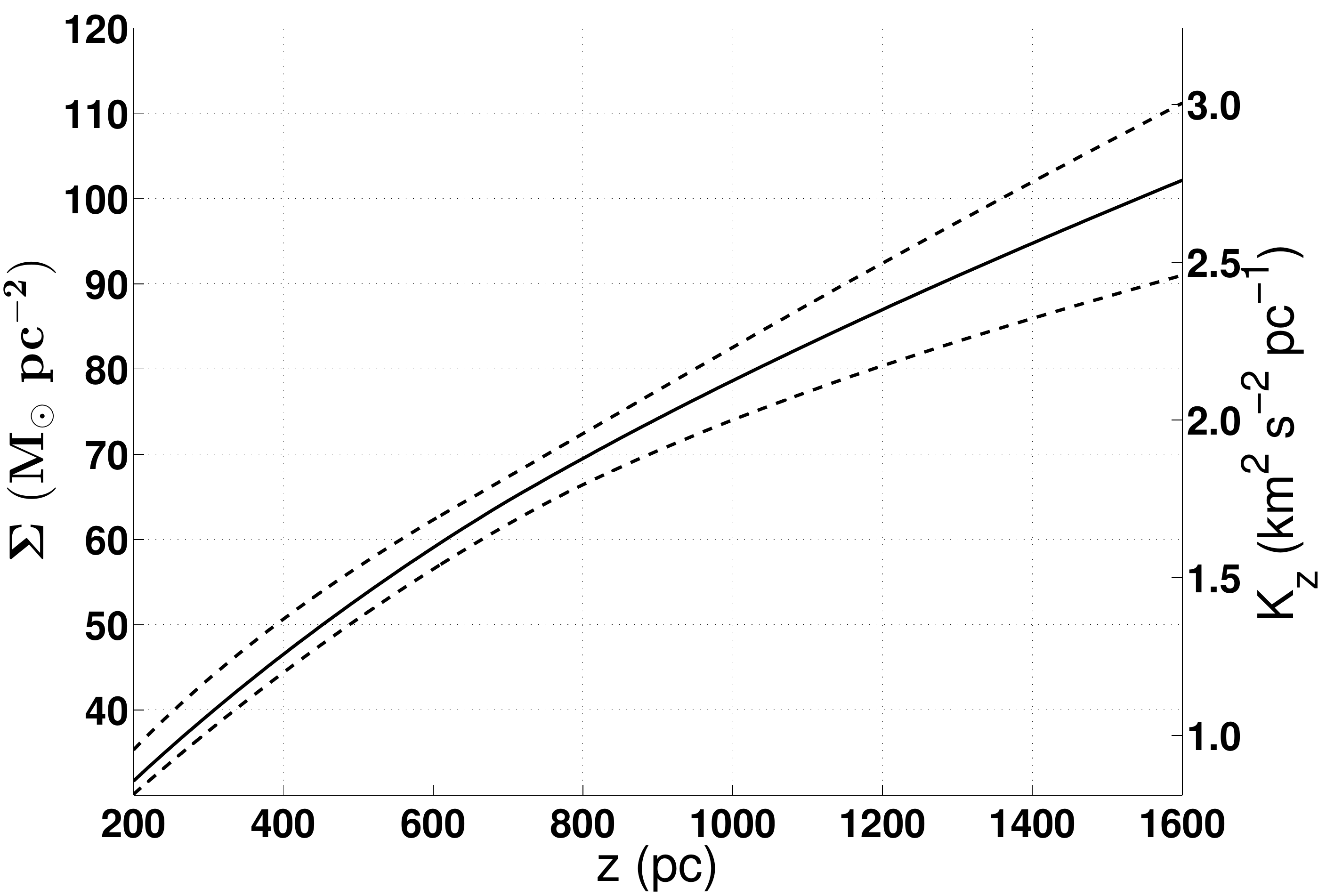} 
 	\caption{The model predicted $K_z(z)$ force, equivalent with the total surface density. The solid line shows the best fit $K_z(z)$, and the two black dashed lines show the 1-$\sigma$ region. The left side y-axis shows the corresponding surface density $\Sigma(z)$.}
	\label{fig:sm}
\end{figure}

The model predicted $K_{z}(z)$ force is shown in Fig.~\ref{fig:sm}. The solid line is the best-fitting $K_{z}(z)$, and the two dashed lines indicate the 1-$\sigma$ region. $K_{z}(z)$ increases with vertical distance $z$ due to the contributions of the stellar disk and the dark matter halo. 
The total surface density up to 1\,kpc can be immediately derived from, 
\begin{equation}\label{eq19}
\Sigma_{\rm tot,\ |z|<1.0\ kpc} = K_{z}(z < 1.0\,{\rm kpc})/{ (- 2\pi G)}.
\end{equation}
We obtain $\Sigma_{\rm tot,\ |z|<1.0\,kpc} = 78.7 ^{+3.9}_{-4.7}\,\rm M_{\odot}\,\rm pc^{-2}$, within which the stellar disk contributes $\Sigma_{\star} = 40.5^{+7.1}_{-6.6}\,\rm M_{\odot}\,\rm pc^{-2}$, and the local dark matter density is $\rho_{\rm  DM} = 0.0159_{-0.0057}^{+0.0047}\,\rm M_{\odot}\,\rm pc^{-3}$. 
If we consider the non-flat nature of the rotation curve, the last term at the right hand side of eq. (\ref{eq6}) is about $-0.0024\,\rm M_\odot\,pc^{-3}$ in the solar neighborhood according to~\citet{BM98}. With this adjustment $\rho_{\rm  DM} = 0.0135_{-0.0057}^{+0.0047}\,\rm M_{\odot}\,\rm pc^{-3}$. 
It is noted that $\Sigma_{\star}$ and $\rho_{\rm DM}$ are anti-correlated, and constrained by the total surface density~\citep{Garbari12}. The bottom-left panel of Fig.~\ref{fig:pdfs} shows this anti-correlation. 



\subsection{Comparison with other works}
The total surface density within 1\,kpc, $\Sigma_{\rm tot,\ |z|<1.0\ kpc} = 78.7 ^{+3.9}_{-4.7} \,\rm M_{\odot}\,\rm pc^{-2}$ obtained here is slightly larger than those from previous works, e.g. \citet{Zhang13} inferred $67 \pm  6\,\rm M_{\odot}\,\rm pc^{-2}$ and \citet{Bienayme14} gave $68.6 \pm 1.0\,\rm M_{\odot}\,\rm pc^{-2}$, but our result is still consistent with theirs within the uncertainties. The local dark matter density, $\rho_{\rm DM}$, estimated here is $0.0159^{+0.0047}_{-0.0057}\,\rm M_{\odot}\,\rm pc^{-3}$
, which is also consistent with previous results. Compared with recent works, our result is larger than \citet{Zhang13} by about 1-$\sigma$, but similar to \citet{Piffl14}, $0.0154 \,\rm M_{\odot}\,\rm pc^{-3}$, and \citet{Bienayme14}, $0.0143\pm 0.0011 \,\rm M_{\odot}\,\rm pc^{-3}$ (see Fig.~\ref{fig:rr}). As we have argued in section~\ref{sect:intro}, a serious comparison should not simply examine the values and their error bars: the large differences in the observed data and models must be taken into account as well.  


\subsection{The uncertainty of the results}
The uncertainty of our result is around 30 percent which is similar to that of \citet{Zhang13} but larger than two RAVE-based works published in 2014 which have an uncertainty of around 10 percent \citep{Piffl14,Bienayme14}. One reason is that the results of their estimates from the Galactic rotation curve (e.g. \citealt{Piffl14}) usually have small error bars due to the strong assumptions about the Galactic halo shape \citep{Read14}, while the local measures usually have larger errors. Another likely reason is due to random noise. If we assume that our model can perfectly describe our solar neighborhood, the uncertainty should only come from sampling noise and observational errors. Below we use a Monte Carlo simulation to quantify the influence of sampling noise. 

\begin{figure}
	\centering
 	\includegraphics[width=3in]{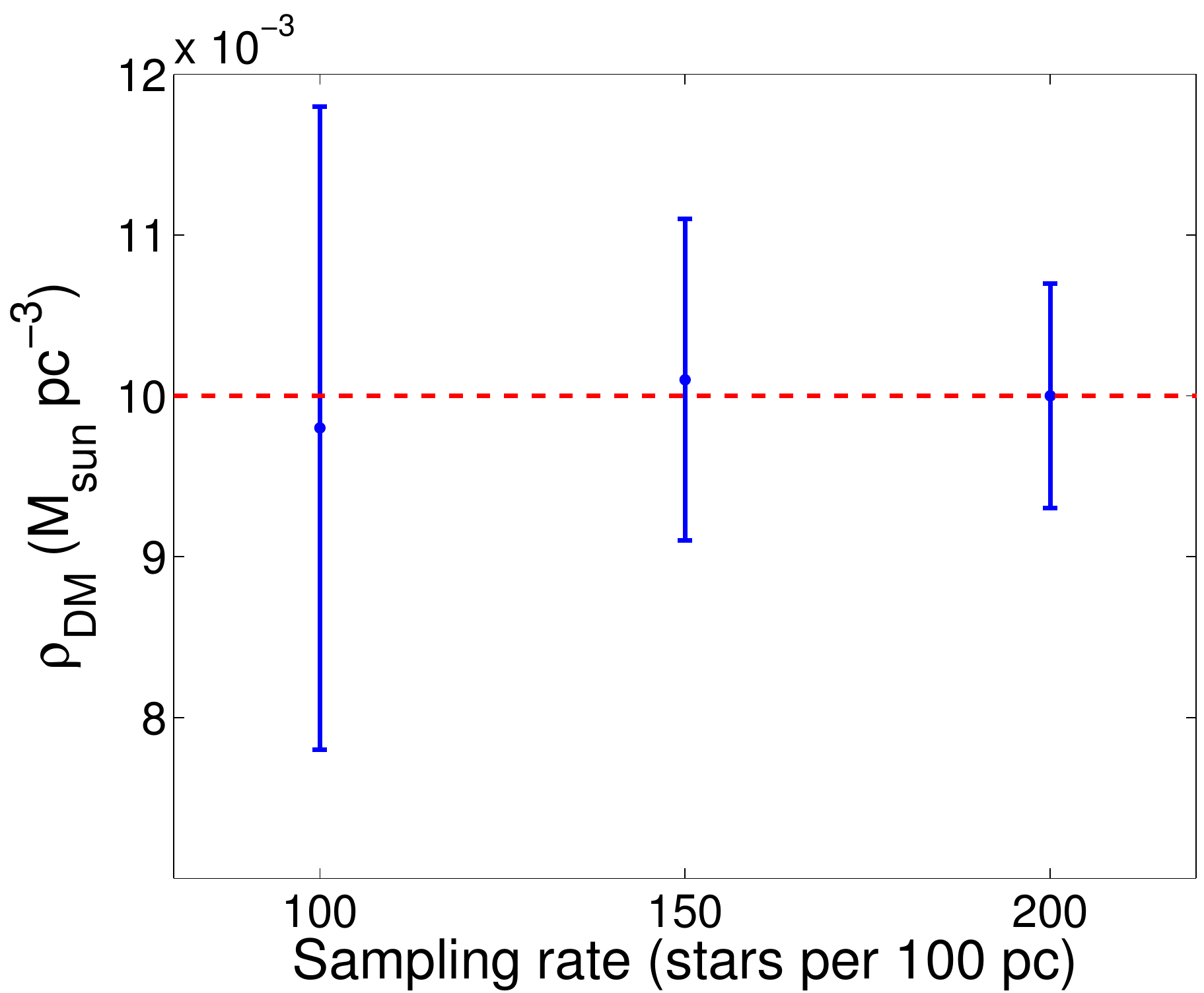} 
 	\caption{The uncertainty caused by the sampling rate. The x-axis is the sampled stars per 100 pc. The y-axis is the averaged dark matter density estimate. The error bars show the 1 $\sigma$ dispersions.}
	\label{fig:SR}
\end{figure}

Using our model, we generate three groups of mock data (all of which assume a local dark matter density of $0.01\,\rm M_{\odot}\,\rm pc^{-3}$) with three sampling densities (i.e. number of stars per unit volume): 1) sampling density of Group 1 is 100 stars per 100 pc, similar to LAMOST DR2; 2) Group 2, 150 stars per 100 pc; 3) Group 3, 200 stars per 100 pc.
Each group contains 100 sets of data. We quantify the influence of random noise by calculating the averaged $\rho_{\rm DM}$ and its variance for the three groups. Fig.~\ref{fig:SR} shows the results from the three groups. The uncertainty is around 20\% of the median value for a sampling density equal to 100 stars per 100 pc, 10\% for 150 stars per 100 pc, and 5\% for 200 stars per 100 pc. It means that with the current LAMOST DR2 sample, the contribution to the uncertainty from the sampling noise is about two-thirds of the total uncertainty which is 30\% based on our model. The remaining one-third of the total uncertainty may come from the measurement errors of velocities, distances, and stellar parameters, or from the systemic errors.

\subsection{The scale height of the stellar mass distribution}
\begin{figure}
	\centering
 	\includegraphics[width=3in]{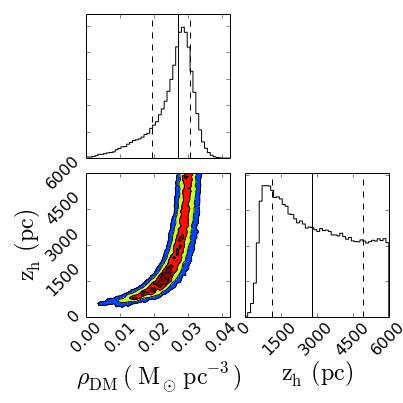} 
 	\caption{The bottom-left panel shows the 2-D PDF of $z_h$ and $\rho_{\rm DM}$ with a much broader prior ($\rm z_h<6000$\,pc). The contours show the 0.5, 1, 1.5, and 2 $\sigma$ levels. The top panel and the bottom-right panel show the 1-D PDF of $\rho_{\rm DM}$ and $z_h$, respectively. The solid line of each panel indicates the median value, while the two dashed lines indicate the 1-$\sigma$ regions.}
	\label{fig:rz6}
\end{figure}

The scale height of the stellar mass distribution, $\rm z_{h}$, is $588_{-192}^{+151}\,\rm pc$, which is not well constrained. In the model fitting procedure a prior $\rm z_{h}<800$\,pc is imposed. It is noted that the \citet{Zhang13} study also suffered from a large uncertainty in this parameter. \citet{Bienayme14} simply fixed $\rm z_{h}$ as 300\,pc. The reason may be that, the Jeans equation method is not sensitive to $\rm z_h$. Unfortunately our sample does not contain enough stars located within $z<300$\,pc to better constrain this quantity. 

As shown in Fig.~\ref{fig:pdfs}, it is clear that $\Sigma_{\star}$ or $\rho_{\rm DM}$ is almost uncorrelated with $\rm z_h$. If we fix $\rm z_h=$ 300\,pc, the local dark matter density, $\rho_{\rm DM}$, is $0.0144 \pm 0.005\,\rm M_{\odot}\,\rm pc^{-3}$, a change of less than 10\%. 
If we relax the prior to $\rm z_h<6000$\,pc, the local dark matter density significantly increases as shown in Fig.~\ref{fig:rz6}. This is because, when $\rm z_h$ is larger than 2000 pc, the contribution of the stellar disk is negligible and the increase in the vertical velocity dispersion depends only on the dark matter. However, such a large $\rm z_h$ seems unrealistic since we know that the dominating stellar component is the thin disk with a scale height of around 300\,pc.

\subsection{Tilt term}\label{sect:tiltterm}
The tilt term in eq. (\ref{eq1}),
\begin{equation}\label{eq20}
\frac{1}{R\ \nu} \frac{\partial}{\partial{R}} (R \nu \sigma^{2}_{Rz}),
\end{equation}
is related to the vertical number density, $\nu(z)$, the cross term of velocity dispersion tensor, $\sigma^{2}_{Rz}$, and their partial derivative with respect to the radius $R$. In most previous works, this term has been neglected (e.g. \citealt{Garbari12,Zhang13}). However,  using SDSS/SEGUE G-type dwarf stars, \citet{Buden15} found that the tilt angle variation with height can be empirically described by the relation
\begin{equation}
{\rm \alpha_{tilt} }= (-0.90 \pm 0.04)\ {\rm arctan}(|z|/R_{0}) - (0.01\pm0.005).
\end{equation}
Here, $\rm \alpha_{tilt}$ is the tilt angle defined as
\begin{equation}
{\rm tan(2\rm \alpha_{tilt})} = \frac{2\sigma^{2}_{Rz}}{\sigma^{2}_{RR}- \sigma^{2}_{zz}}.
\end{equation}
They concluded that, the tilt term cannot be ignored. \citet{Sil15} subsequently modelled the tilt term and investigated its influence using simulation data. They found that the contribution to the tilt term by thick disk stars is about 9\% in $\rm K_{z}$ at z = 1.5\,kpc, while the contribution from the thin disk stars is only about 5\% at the same height (see their Fig. 2). Since we use only the thin disk stars in this work and the maximum height in our sample is 1.5\,kpc, the influence of the tilt term can be neglected if the \citet{Sil15} findings are accepted. To confirm this, we run a modified model taking into account the tilt term modelled by \citet{Sil15} and found a local dark matter density of $0.0153^{+0.0048}_{-0.0057}\,\rm M_{\odot}\,\rm pc^{-3}$, which is consistent with our previous result.

\subsection{A non-equilibrium component of the disk}
A fundamental assumption of the measurement of the local dark matter density is that the tracers have to be individually in equilibrium so that the collisionless Boltzmann equation-based models can be applied. However, recent observations found some evidence of asymmetric vertical motions in the Galactic disk, which violate the equilibrium assumption (\citealt{W13,Carlin13,Xu15}). They are likely wave modes induced by vertical perturbations due to dwarf satellites, the triaxial dark matter halo, or spiral structures. Theoretical works show that about 20\% of stars in the disk may be perturbed by dwarf satellites and the mean vertical velocity can be shifted by 5--10 $\rm km\ s^{-1}$ \citep{Gomez13}. Using LAMOST data, \citet{Carlin13} also found that the mean vertical velocity is shifted by at most 10 $\rm km\ s^{-1}$. This challenges the methodology of the local dark matter density determination, not only because of the non-equilibrium stars themselves but also because of their effects on other stars which may lead to systematic errors.

\begin{figure}
	\centering
 	\includegraphics[scale=0.291]{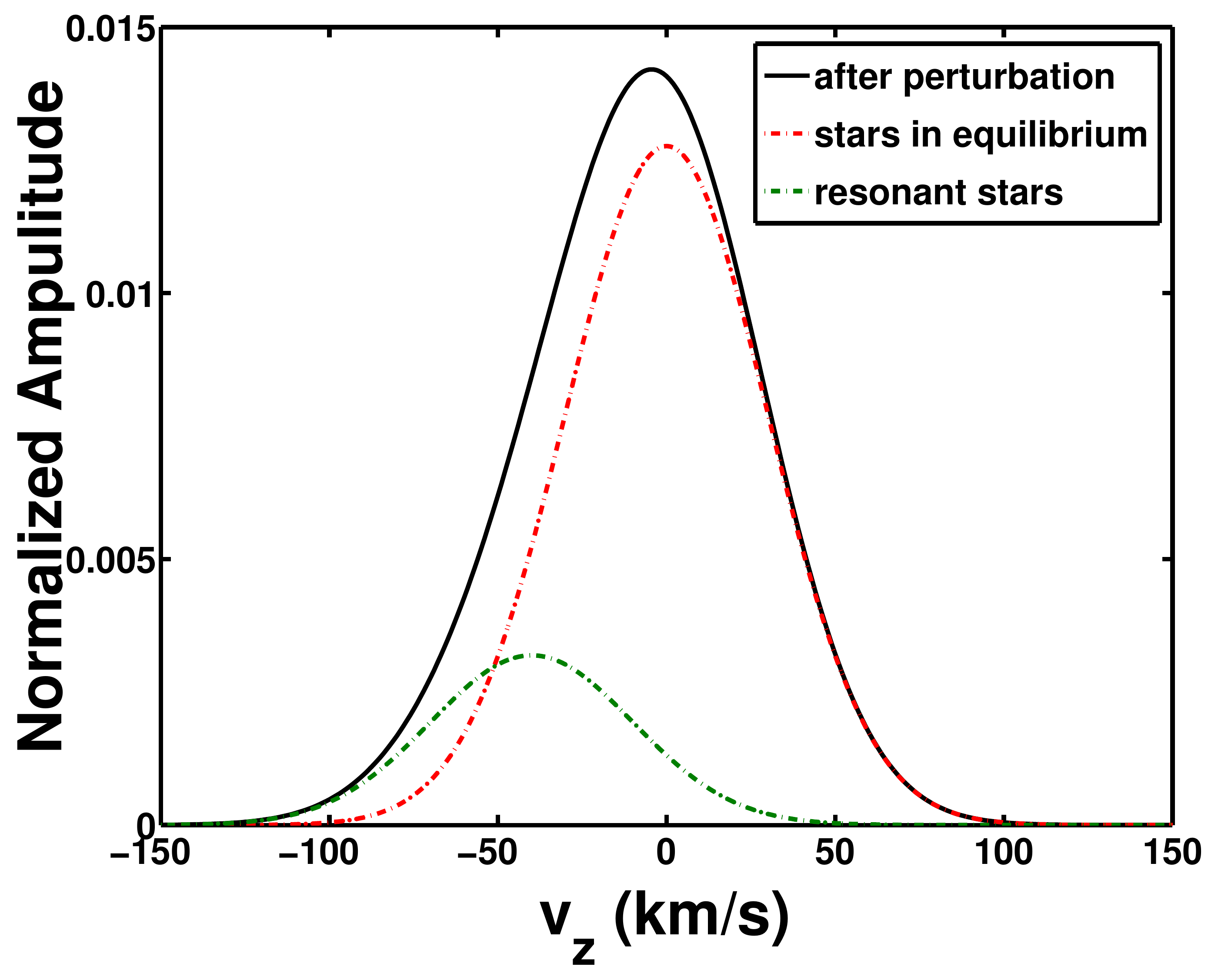}
 	\includegraphics[scale=0.291]{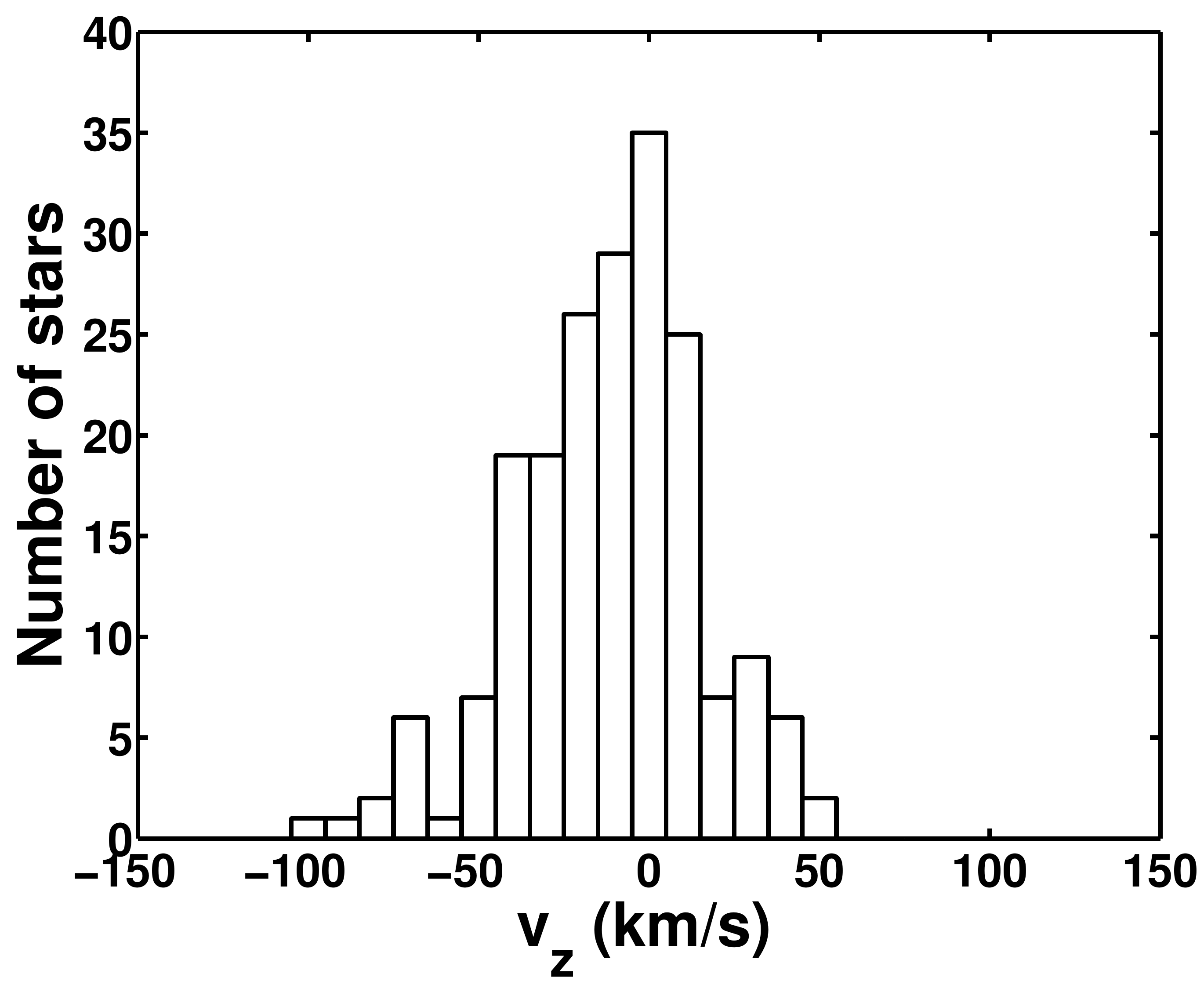}
 	\caption{The top panel shows the mock data with an additional asymmetric velocity component to mimic a perturbation. The initial mean velocity is 0 $\rm km\ s^{-1}$ and the velocity dispersion is 30\,$\rm km\,s^{-1}$. We assumed that 20\% (the blue dashed line) of the stars shift their velocities by 40\,$\rm km\,s^{-1}$ induced by perturbation and the rest (the red dashed line) are still in equilibrium. The resulting mean velocity of the total sample (the black solid line) has a shift of about 8 $\rm km\ s^{-1}$ and the velocity dispersion increases by about 4 $\rm km\ s^{-1}$. The bottom panel shows the vertical velocity distribution of the LAMOST sample located at the vertical distance between 0.5\,kpc and 0.6\,kpc.}
	\label{fig:shift}
\end{figure}

We run a simple test to evaluate this effect. We give an additional vertical velocity of 40\,$\rm km\,s^{-1}$ for a subset of 20\% of randomly selected stars to mimic a perturbation. The vertical velocity distribution is no longer symmetric, as shown by the black line in the top panel of Fig.~\ref{fig:shift}. The derived velocity dispersion without considering the non-equilibrium component increases by about 4\,$\rm km\,s^{-1}$ with a shift of the mean velocity by about 8\,$\rm km\,s^{-1}$. This would lead to an overestimation of the local dark matter density by approximately 40\%. The observed velocity distribution of the LAMOST sample between 0.5\,kpc and 0.6\,kpc, as used in this work, is shown in the bottom panel of Fig.~\ref{fig:shift}. Although it is hard to identify an additional velocity component, it does show a slight asymmetry. In future works, we plan to address this issue by combining  about 1000 stars toward the South Galactic Pole with the LAMOST sample such that the wave modes in the vertical velocity can be identified and fully considered in an improved measurement of the local dark matter density.

\subsection{The thick disk contaminations}\label{sect:thickdisc}
In section~\ref{sect:data} we use [Fe/H]$>-0.5$\,dex to remove the thick disk. In this section we test how many thick disk stars still remain after this cut and whether they affect the measurement of the $K_z$ force.
The potential contamination from the thick disk mostly affects the high $z$ regime. For G type stars (roughly corresponding to $5300<\rm T_{eff}<6000$\,K), the thick disk contamination under this criterion is $\sim33$\% at $600<z<1500$\,pc according to the data in~\citet{Liu12} with a similar cut in volume\footnotemark[2]\footnotetext[2]{The thick disk is defined as [$\alpha$/Fe]$>0.25$\,dex. The data is selected with $b>70^\circ$ instead of $85^\circ$ to ensure sufficient stars in the statistics.}. For K type stars (roughly corresponding to $4000<\rm T_{eff}<5300$\,K), according to Fig. 1 in~\citet{Zhang13}, the thick disk contamination (defined as [$\alpha$/Fe]$>0.25$\,dex) is also quite low, similar to that for G type stars. Considering the mixture of the thick disk stars, the scale height of our sample is 10\% larger than the scale height of the thin disk. 

We further perform a simple test to estimate how significant these contaminations would be in the velocity dispersion. According to~\citet{Zhang13}, the thin disk (roughly corresponding to their group I with a scale height of $\sim$ 260\,pc) has $\sigma_z\sim21$\,kms$^{-1}$ at $600<z<1500$\,pc and possibly contaminating thick disk stars (mostly from their group II with scale height $\sim$ 450\,pc) have $\sigma_z\sim28$\,kms$^{-1}$. If the 33\% stars are from the thick disk contamination, then the derived velocity dispersion with contamination would be about 3\,km\,s$^{-1}$ larger than that of the thin disk. 
In this case, we will overestimate the local dark matter density $\rho_{DM}$ by about 20\%, and the value of $\rho_{\rm DM}$ becomes $0.0133\,\rm M_{\odot}\,\rm pc^{-3}$. Future data releases of LAMOST will contain information on [$\alpha$/Fe], which will reduce the contamination of thick disk stars significantly.

\section{Conclusion}\label{sect:conclusion}

Determination of the local dark matter density from the vertical motions of the stars is an important application in Galactic dynamics; it is also an important input parameter in determining the dark matter flux in direct detection experiments. However, in the past century, the estimated values by many works are far from being in agreement with each other. Either the dynamical models and/or the observed data significantly affect the measured values. In this work, we build a well defined observational sample based on the LAMOST spectroscopic survey. Uncertainties in stellar parameters, distances, line-of-sight velocities, selection effects, volume completeness, and stellar populations are all carefully taken into account. We finally obtain a sample with 1427 G \& K type main-sequence stars, most of which belong to the thin disk population following an exponential vertical density profile. Based on this sample we execute a series of modelling works to estimate the local dark matter density. We start with the vertical Jeans equation method and assume a single exponential stellar mass distribution, a razor thin gas disk, and a constant dark matter distribution within $1.5$\,kpc. We estimate the posterior probability density function for the model parameters, including the surface mass density of the disk and the local volume mass density of the dark matter, with Markov Chain Monte Carlo simulations. We obtain a total surface mass density of $78.7 ^{+3.9}_{-4.7} \,\rm M_{\odot}\,\rm pc^{-2}$ up to 1\,kpc and a local dark matter density of $0.0159^{+0.0047}_{-0.0057}\,\rm M_{\odot}\,\rm pc^{-3}$ if the rotation curve is flat. 

Although these results are consistent with most recent works, the agreements may be fortuitous because neither the methods nor the data are the same. Even relative uncertainties are not comparable because oversimplified assumptions can lead to unrealistically smaller errors. We found that for our case the sampling noise of the observed data contributes about two-thirds of the uncertainty in our determination of the local dark matter density.  The tilt term, a non-equilibrium component induced by perturbations and contaminations from the thick disk for example, can explain the remaining uncertainty. 

In future works, we plan to combine LAMOST data for the observed northern Galactic pole with new data from the southern Galactic pole to identify better any non-equilibrium components and to improve estimation of both the vertical stellar density and the velocity dispersion profile. This augmented sample will enhance our ability to evaluate different assumptions within dynamical models.  In the end, this will be beneficial in improving local dark matter density determinations.
   

\section*{Acknowledgment}
This work is supported by the Strategic Priority Research Program ``The Emergence of Cosmological Structures" of the Chinese Academy of Sciences, Grant No. XDB09000000, the National Key Basic Research Program of China 2014CB845700, and the National Natural Science Foundation of China (NSFC) under grant 11333003 and 11390372. CL acknowledges the NSFC under grants 11373032 and U1231119. Guoshoujing Telescope (the Large Sky Area Multi-Object Fiber Spectroscopic Telescope LAMOST) is a National Major Scientific Project built by the Chinese Academy of Sciences. Funding for the project has been provided by the National Development and Reform Commission. LAMOST is operated and managed by the National Astronomical Observatories, Chinese Academy of Sciences.



\clearpage

\begin{thebibliography}{}

\bibitem[{{An} et al. (2009)}]{An09} An, D., Pinsonneault, M. H., Masseron, T., et al.\ 2009, ApJ, 700, 523

\bibitem[{{Bahcall} (1984a)}]{Bahcall84a} Bahcall, J. N.,\ 1984a, ApJ, 287, 926

\bibitem[{{Bahcall} (1984b)}]{Bahcall84b} Bahcall, J. N.,\ 1984b, ApJ, 276, 169

\bibitem[{{Bahcall} et al. (1992)}]{Bahcall92} Bahcall, J. N., Flynn C., Gould A.,\ 1992, ApJ, 389, 234

\bibitem[{{Bienayme} et al. (1987)}]{Bienayme87} Bienayme, O., Robin, A. C., Creze, M.,\ 1987, A\&A, 180, 94

\bibitem[{{Bienaym{\'e} et al.} (2014)}]{Bienayme06} Bienaym{\'e}, O., Soubiran, C., Mishenina, T. V., Kovtyukh, V. V., Siebert, A.,\ 2006, A\&A, 446, 933

\bibitem[{{Bienaym{\'e} et al.} (2014)}]{Bienayme14} Bienaym{\'e}, O., Famaey, B., Siebert, A., et al.\ 2014, A\&A, 571, 92

\bibitem[\protect\citeauthoryear{Binney 
\& Merrifield}{1998}]{BM98} Binney, J., Merrifield, M., Galactic Astronomy, Princeton, NJ : Princeton University Press, 1998 

\bibitem[{{Binney} \& {Tremaine} (2008)}]{BT08} Binney, J. \& Tremaine, S., 2008, Galactic Dynamics: Second Edition, Princeton University Press

\bibitem[{{Bovy} et al. (2012a)}]{B12a}Bovy, J., Allende Prieto, C., Beers, T. C., et al. 2012a, ApJ, 759, 131

\bibitem[{{Bovy} \& {Tremaine} (2012b)}]{BT12} Bovy, J., Tremaine, S.,\ 2012b, ApJ, 756, 89

\bibitem[{{Bovy} \& {Rix} (2013)}]{BR13} Bovy, J., Rix, H.-W.,\ 2013, ApJ, 779, 115

\bibitem[{{B{\"u}denbender} et al. (2015)}]{Buden15} B{\"u}denbender, A., van de Ven, G., Watkins, L. L.,\ 2015, MNRAS, 452, 956

\bibitem[\protect\citeauthoryear{Carlin et al.}{2012}]{Carlin12} 
Carlin, J.~L., et al., 2012, RAA, 12, 755 

\bibitem[{{Carlin} et al. (2013)}]{Carlin13} Carlin, J. L., DeLaunay, J., Newberg, H. J., et al.\ 2013, ApJ, 777, 5

\bibitem[{{Carlin} et al. (2015)}]{Carlin15} Carlin, J. L., Liu, C., Newberg, H., et al.\ 2015, AJ, 150, 4

\bibitem[{{Catena} \& {Ullio} (2010)}]{CU10} Catena, R., Ullio, P.,\ 2010, JCAP, 8, 4

\bibitem[{{Creze} et al. (1998)}]{Creze98} Creze, M., Chereul, E., Bienayme, O., Pichon, C.,\ 1998, A\&A, 329, 920

\bibitem[{{Cui} et al. (2012)}]{Cui12} Cui, X.~Q., Zhao, Y.~H., Chu, Y.~Q., Li, G.~P., et al.\ 2012, RAA, 12, 1197


\bibitem[{{Deng} et al. (2012)}]{Deng12} Deng, L.~C., Newberg, H., Liu, C., et al.\ 2012, RAA, 12, 735

\bibitem[{{Dehnen} \& {Binney} (1998)}]{DB98} Dehnen, W., Binney, J.,\ 1998, MNRAS, 294, 429

\bibitem[{{Fich} et al. (1989)}]{Fich89} Fich, M., Blitz, L., Stark, A. A.,\ 1989, ApJ, 342, 272

\bibitem[{{Flynn} et al. (2006)}]{Flynn06} Flynn, C., Holmberg, J., Portinari, L., Fuchs, B., Jahrei{\ss}, H.\ 2006, MNRAS, 372, 1149

\bibitem[{{Gao} et al. (2014)}]{Gao14} Gao, S., Liu, C., Zhang, X.~B., et al.\ 2014, ApJ, 788, 37

\bibitem[{{Garbari} et al. (2011)}]{Garbari11} Garbari, S., Read, J. I., Lake, G.,\ 2011, MNRAS, 416, 2318

\bibitem[{{Garbari} et al. (2012)}]{Garbari12} Garbari, S., Liu, C., Read, J. I., Lake, G., 2012, MNRAS, 425, 1445

\bibitem[{{G{\'o}mez} et al. (2013)}]{Gomez13} G{\'o}mez, F. A., Minchev, I., O'Shea, B. W., et al. 2013, MNRAS, 429, 159


\bibitem[\protect\citeauthoryear{Hessman}{2015}]{Hessman15} Hessman, F.~V., 2015, A\&A, 579, A123

\bibitem[{{Hill} (1960)}]{Hill60} Hill, E. R.,\ 1960, Bull. Astron. Inst. Netherlands, 15

\bibitem[{{Holmberg} \& {Flynn} (2000)}]{HF00} Holmberg, J., Flynn, C.,\ 2000, MNRAS, 313, 209

\bibitem[{{Holmberg} \& {Flynn} (2004)}]{HF04} Holmberg, J., Flynn, C.,\ 2004, MNRAS, 352, 440

\bibitem[{{Ivezi{\'e}} et al. (2008)}]{Ivezic08} Ivezi{\'c}, \u{Z}., Sesar, B., Juri{\'c}, M., et al.\ 2008, ApJ, 684, 287

\bibitem[\protect\citeauthoryear{Juri{\'c} et 
al.}{2008}]{Juric08} Juri{\'c}, M., et al., 2008, ApJ, 673, 864

\bibitem[{{Kapteyn} (1922)}]{Kapteyn22} Kapteyn, J. C., 1922,\ ApJ, 55, 302

\bibitem[{{Kuijken} \& {Gilmore} (1989a)}]{KG89a} Kuijken, K., Gilmore, G.,\ 1989a, MNRAS, 239, 571

\bibitem[{{Kuijken} \& {Gilmore} (1989b)}]{KG89b} Kuijken, K., Gilmore, G.,\ 1989b, MNRAS, 239, 605

\bibitem[{{Kuijken} \& {Gilmore} (1989c)}]{KG89c} Kuijken, K., Gilmore, G.,\ 1989c, MNRAS, 239, 651

\bibitem[{{Kuijken} \& {Gilmore} (1991)}]{KG91} Kuijken, K., Gilmore, G.,\ 1991, ApJ, 367, L9

\bibitem[\protect\citeauthoryear{Liu 
\& van de Ven}{2012}]{Liu12} Liu, C., van de Ven, G., 2012, MNRAS, 425, 2144

\bibitem[{{McKee} et al. (2015)}]{McKee15} McKee, C. F., Parravano, A., Hollenbach, D. J.,\ 2015, arXiv:1509.05334

\bibitem[{{McMillan} (2011)}]{McMillan11} McMillan, P. J.,\ 2011, MNRAS, 414, 2446

\bibitem[{{Merrifield} (1992)}]{Merrifield92} Merrifield, M. R.,\ 1992, AJ, 103, 1552

\bibitem[{{Moni Bidin} et al. (2012)}]{MB12} Moni Bidin, C., Carraro, G., M{\'e}ndez, R. A., Smith, R.,\ 2012, ApJ, 751, 30

\bibitem[{{Oort} (1932)}]{Oort32} Oort, J. H.,\ 1932, Bulletin of the Astronomical Institutes of the Netherlands, 6, 249

\bibitem[{{Oort} (1960)}]{Oort60} Oort, J. H.,\ 1960, Bulletin of the Astronomical Institutes of the Netherlands, 15, 45

\bibitem[{{Piffl} et al. (2014)}]{Piffl14} Piffle, T., Binney, J., McMillan, P. J., et al.\ 2014, MNRAS, 445, 3133

\bibitem[{{Read} (2014)}]{Read14} Read, J. I.,\ 2014, arXiv:1404.1938

\bibitem[{{Siebert} et al. (2003)}]{Siebert03} Siebert, A., Bienaym{\'e}, O., Soubiran, C.,\ 2003, A\&A, 399, 531

\bibitem[{{Silverwood} et al. (2015)}]{Sil15} Silverwood, H., Sivertsson, S., Steger, P., Read, J. I., Bertone, G.,\ 2015, arXiv:1507.08581

\bibitem[{{Smith} et al. (2012)}]{Smith12} Smith, M. C., Whiteoak, S. H., Evans, N. W.,\ 2012, ApJ, 746, 181

\bibitem[{{Sofue} et al. (2009)}]{Sofue09} Sofue, Y., Honma, M., Omodaka, T.,\ 2009, PASJ, 61, 227

\bibitem[\protect\citeauthoryear{Skrutskie et 
al.}{2006}]{2mass} Skrutskie, M.~F., et al., 2006, AJ, 131, 
1163 

\bibitem[\protect\citeauthoryear{Tian et al.}{2015}]{tian15} 
Tian, H.-J., et al., 2015, ApJ, 809, 145 

\bibitem[{{Weber} \& {de Boer} (2010)}]{WB10} Weber, M., de Boer, W.,\ 2010, A\&A, 509, A25


\bibitem[{{Williams} {et~al.}(2013){Williams}}]{W13}
 {Williams}, M. E. K., et al. 2013, MNRAS, 436, 101

\bibitem[{Wu} et al. (2014)]{Wu14} Wu, Y., Du, B., Luo, A., et al., 2014, IAUS, 306, 340

\bibitem[{{Xu} et al. (2015)}]{Xu15} Xu, Y., Newberg, H. J., Carlin, J. L., et al. 2015, ApJ, 801, 105

\bibitem[{{Zhang} et al. (2013)}]{Zhang13} Zhang, L., Rix, H.-W., van de Ven, G., Bovy, J., Liu, C., Zhao, G.,\ 2013, ApJ, 772, 108

\bibitem[{{Zhao} et al. (2012)}]{Zhao12} Zhao, G., Zhao, Y.~H., Chu, Y.~Q., Jing, Y.~P., Deng, L.~C.\ 2012, RAA, 12, 723

\end{thebibliography}
\end{document}